SRNR: Training neural networks for <u>S</u>uper-<u>R</u>esolution MRI using <u>N</u>oisy high-resolution <u>R</u>eference data

Jiaxin Xiao[1], Zihan Li[2], Berkin Bilgic[3,4], Jonathan R. Polimeni[3,4], Susie Huang[3,4], Qiyuan Tian[3,4]

[1]Department of Electronic Engineering, Tsinghua University, Beijing, P.R. China;
[2]Department of Biomedical Engineering, Tsinghua University, Beijing, P.R. China;
[3]Athinoula A. Martinos Center for Biomedical Imaging, Department of Radiology, Massachusetts General Hospital, Charlestown, MA, USA;
[4]Harvard Medical School, Boston, MA, USA;

**Primary label**
Machine Learning/Artificial Intelligence

**Secondary label**
Data Processing

**Summary of Main Findings**
- Noise in the high-resolution reference data only slightly compromises super-resolution performance of neural networks.
- The number of repetitions of reference data for averaging can be reduced to shorten acquisition time and increase the feasibility and accessibility.

**Synopsis**

Neural network (NN) based approaches for super-resolution MRI typically require high-SNR high-resolution reference data acquired in many subjects, which is time consuming and a barrier to feasible and accessible implementation. We propose to train NNs for Super-Resolution using Noisy Reference data (SRNR), leveraging the mechanism of the classic NN-based denoising method Noise2Noise. We systematically demonstrate that results from NNs trained using noisy and high-SNR references are similar for both simulated and empirical data. SRNR suggests a smaller number of repetitions of high-resolution reference data can be used to simplify the training data preparation for super-resolution MRI.

**Introduction**

Super-resolution MRI synthesizes high-resolution images from low-resolution images and is a useful tool to shorten the scan time and visualize fine structures. Deep learning techniques, particularly neural networks (NNs), have been shown to outperform conventional super-resolution methods and are widely adopted in biomedical imaging[1-4]. Nonetheless, the training of NNs requires high-quality and high-resolution reference data acquired in many subjects, as a result reducing the practical feasibility and accessibility of NN-based super-resolution. Higher resolution necessitates not only a larger number of data samples and longer acquisition times to encode a whole-brain volume, but also multiple signal averages to improve the inherently low signal-to-noise ratio (SNR). This work aims to simplify reference data acquisition for NN-based super-resolution.

Noise2Noise[5] is a classic NN-based denoising method that does not need high-SNR reference data, which works by training denoising NNs to map a noisy image to another instance of the image acquired with a statistically independent noise sample. It is proven that the learned NN parameters remain unchanged if certain statistics of noisy reference values match those of the clean reference values (e.g., the expectation for L2 minimization).

**In this work, we leverage the mechanism of Noise2Noise and train NNs for Super-Resolution using Noisy Reference data (SRNR).** The proposed SRNR method trains NNs to map low-resolution images to their residuals compared to noisy high-resolution reference images, which are composed of random noise and static high-frequency residuals resulting from the difference in resolution. We hypothesize that the random noise does not affect the training.

We systematically demonstrate the efficacy of SRNR by comparing its results to those from NNs trained using high-SNR references. We first evaluate the effects of noise in reference data on super-resolved images using simulated reference data with different levels of noise. We then show that super-resolution results from NNs trained with single-average and 10-average high-resolution data are highly similar. SRNR has the potential to promote wider adoption of NN-based super-resolution MRI, thereby benefiting a broader range of clinical and neuroscientific applications.

**Methods**

Simulated data. High-SNR T1w images acquired at $0.7 \times 0.7 \times 0.7$ mm$^3$ resolution of 30 healthy subjects from the Human Connectome Project (HCP)[6, 7] WU-Minn-Ox Consortium were used as the ground-truth high-resolution data. Low-resolution thick-slice images were generated by down-sampling the ground-truth data along the superior-inferior direction to $0.7 \times 0.7 \times 3.5$ mm$^3$ resolution. Noisy high-resolution data were simulated by adding Gaussian noise (μ=0, σ=0.2, 0.4, 0.6, …, 1.6×σ of brain voxel intensities) to the native data as the noisy reference high-resolution data (Fig.1b).

Empirical data. Highly accelerated (R=3×3) high-resolution ($0.8 \times 0.8 \times 0.8$ mm$^3$, 10 repetitions) and lower-resolution ($1 \times 1 \times 1$ mm$^3$, 1 repetition) T1w data were acquired on 9 healthy subjects using a Wave-MPRAGE sequence (five on a MAGNETOM Skyra scanner and four on a MAGNETOM Prisma scanner,

Siemens Healthineers). For each subject, the 10 repetitions of high-resolution T1w data were co-registered to the first repetition and averaged using FreeSurfer's "*mri_robust_template*" function[8-11]. The low-resolution T1w data were co-registered to the 10-repetition averaged volume using NiftyReg's "reg_aladin" function.

Network and training. The modified U-Net (MU-Net, Fig.1a) with all pooling and up-sampling layers removed was adopted and implemented using Tensorflow. The input of MU-Net was the low-resolution image volume up-sampled to the target high-resolution using cubic spline interpolation. The output of MU-Net was the residual volume between the input and target volume. The training and validation were performed on the simulation data of 15 subjects with the high-SNR and noisy high-resolution data as the target as well as the empirical data of 4 subjects with the single-average and 10-average data as the target respectively, with Adam optimizers and L2 loss.

Evaluation. The mean absolute error (MAE), peak SNR (PSNR), and structural similarity index (SSIM) were computed to quantify the similarity between up-sampled, super-resolved images and ground-truth high-resolution images, as well as between super-resolved images from different models.

**Results**

Super-resolved images from NNs trained with noisy high-resolution references were visually (Figs.2 and.4) and quantitatively (Figs.3 and 5) similar to those from NNs trained with high-SNR reference data. For the simulated data, the similarity gradually decreased as the noise level in the reference data increased, as expected (Fig.3). Even for the noisiest case (Fig.2, vii), the similarity compared to the ground-truth data only slightly decreased (Fig.3d, MAE: 0.0214 vs. 0.0238, 11.2%; PSNR: 30.26 dB vs. 29.43 dB, -2.7%; SSIM: 0.906 vs. 0.890, -1.77%) and **the inter-result similarity was high** (Fig.3e, MAE: 0.0122; PSNR: 34.61 dB; SSIM: 0.972).

For the empirical data, the similarity of results using single-average and 10-average reference data compared to the ground-truth data was highly similar (Fig.5d, MAE: 0.0203 vs. 0.0215, 5.9%; PSNR: 30.90 dB vs. 30.44 dB, -1.6%; SSIM: 0.930 vs. 0.925, -0.54%). The inter-result similarity was also high (MAE: 0.0097; PSNR: 36.75 dB; SSIM: 0.988).

**Discussion and Conclusion**
**SRNR shows comparable efficacy of training using noisy versus high-SNR high-resolution reference data for super-resolution MRI.** The success of the SRNR approach suggests that **a smaller number of repetitions of high-resolution reference data for averaging can be adopted** to achieve slightly compromised super-resolution performance and improve feasibility and accessibility. Using a single repetition also avoids additional data re-sampling during image co-registration, thus avoiding potential blurring of the high-resolution reference data.

**Acknowledgments**

The T$_1$-weighted data for simulation was provided by the Human Connectome Project, WU-Minn-Ox Consortium (Principal Investigators: David Van Essen and Kamil Ugurbil; U54-MH091657) funded by the 16 NIH Institutes and Centers that support the NIH Blueprint for Neuroscience Research; and by the McDonnell Center for Systems Neuroscience at Washington University. This work was supported by the National Institutes of Health (grant numbers P41-EB015896, P41-EB030006, K99-AG073506), and the Athinoula A. Martinos Center for Biomedical Imaging.

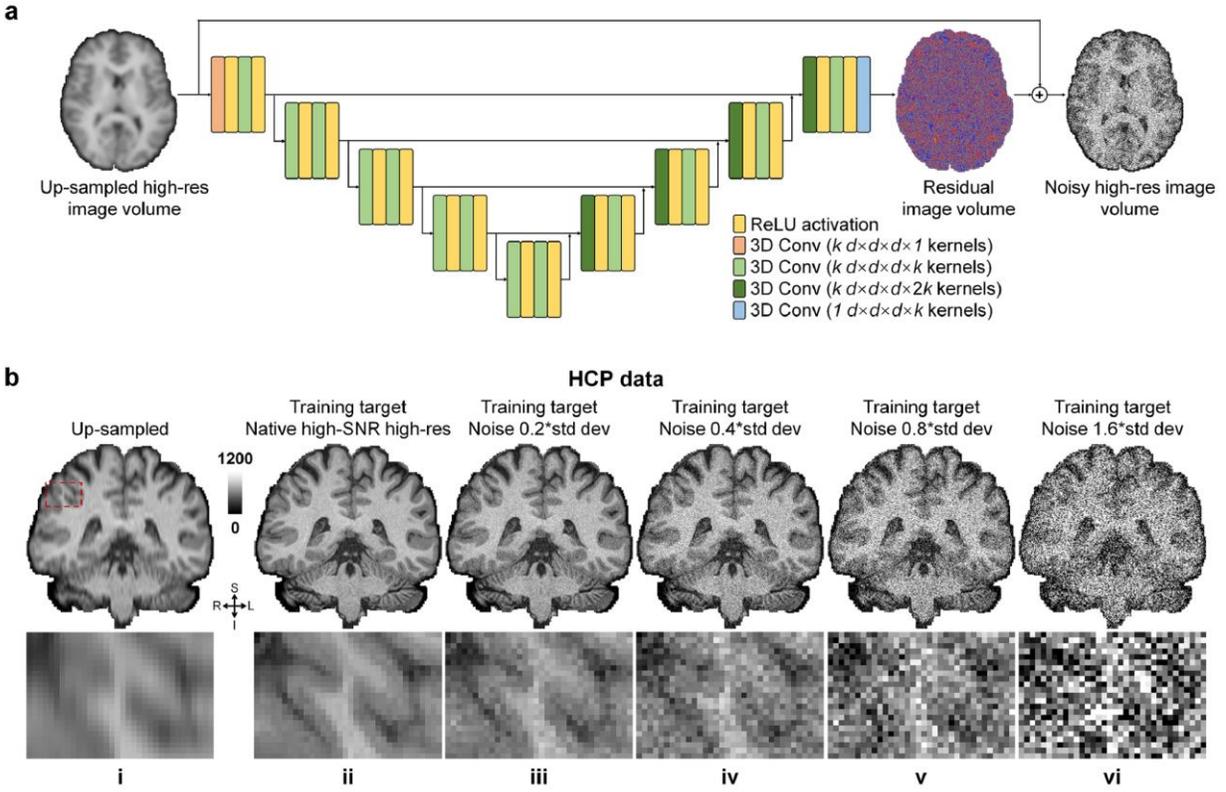

**Figure 1. Neural network and training data.** SRNR uses a modified 3D U-Net (MU-Net) composed of 3D convolution layers and ReLU activation layers with skip connection. MU-Net maps the input up-sampled low-resolution image volume to the residual volume between input and output noisy high-resolution image volume (a). Exemplary coronal image slices from up-sampled low-resolution (b, i), high-SNR high-resolution (b, ii), and simulated noisy high-resolution images (b, ii-vi, with different standard deviations of noise) of a representative HCP subject are shown with enlarged views.

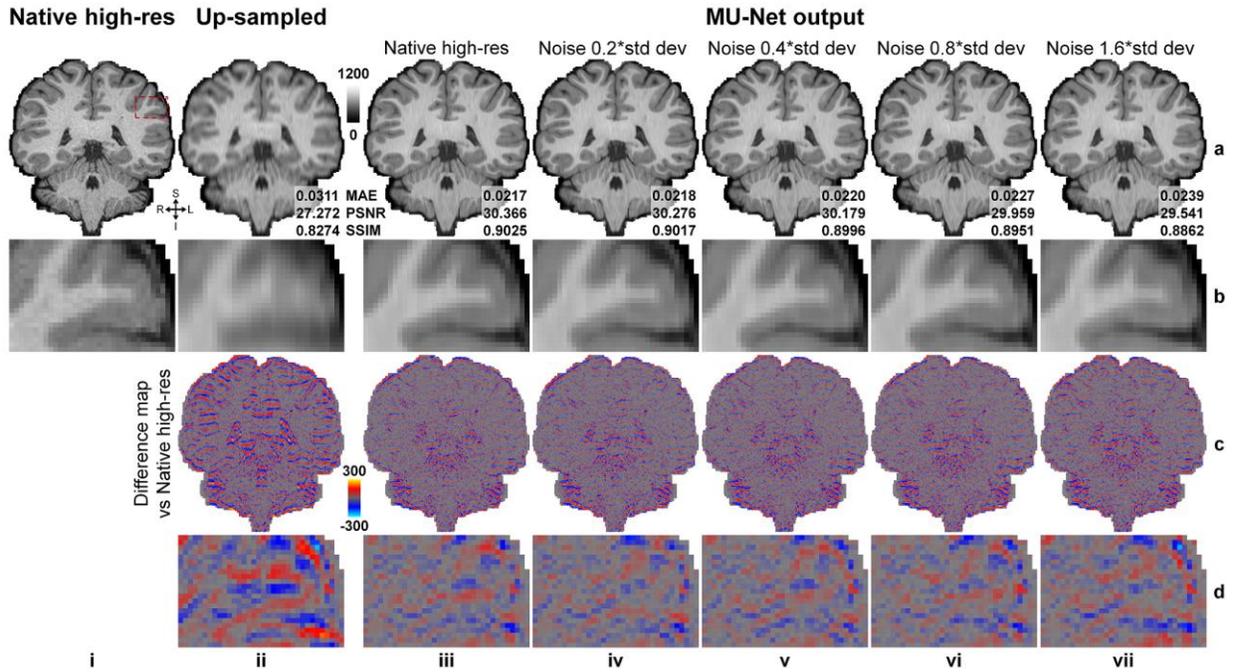

**Figure 2. Simulation image results.** Exemplary coronal image slices from native high-SNR high-resolution (a, i) and up-sampled low-resolution data (a, ii), and super-resolution results from neural networks trained with noisy high-resolution images (a, iii-vii, with different standard deviations of noise) are shown with enlarged views (b). The difference maps (c, d) depict the similarity compared to the native high-SNR high-resolution image. Mean absolute error (MAE), peak SNR (PSNR), and structural similarity index (SSIM) are listed to quantify the similarity (a).

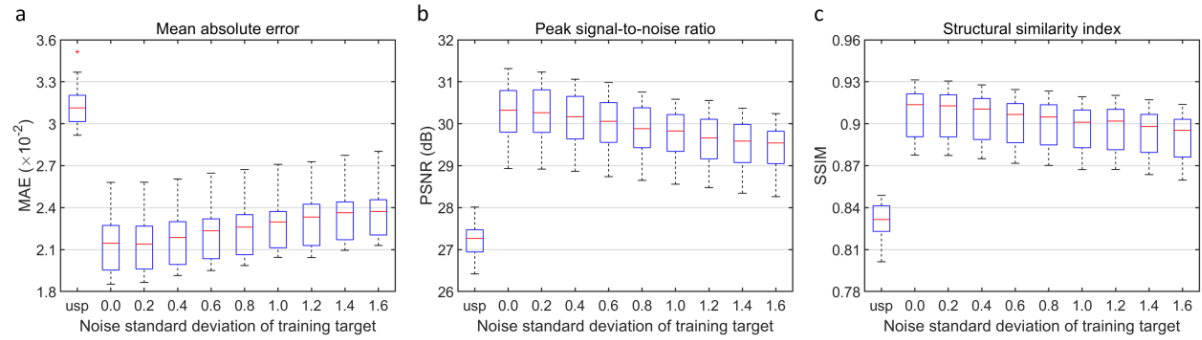

**Figure 3. Simulation image similarity metrics.** Group mean (± standard deviation) of the whole-brain averaged mean absolute error (MAE) (a), peak SNR (PSNR) (b), and structural similarity index (SSIM) (c) of the up-sampled low-resolution images (usp) and results from neural networks trained with high-SNR and noisy high-resolution images across 15 evaluation subjects. Values compared to native high-resolution are listed in table (d), compared to high-SNR training results are listed in table (e).

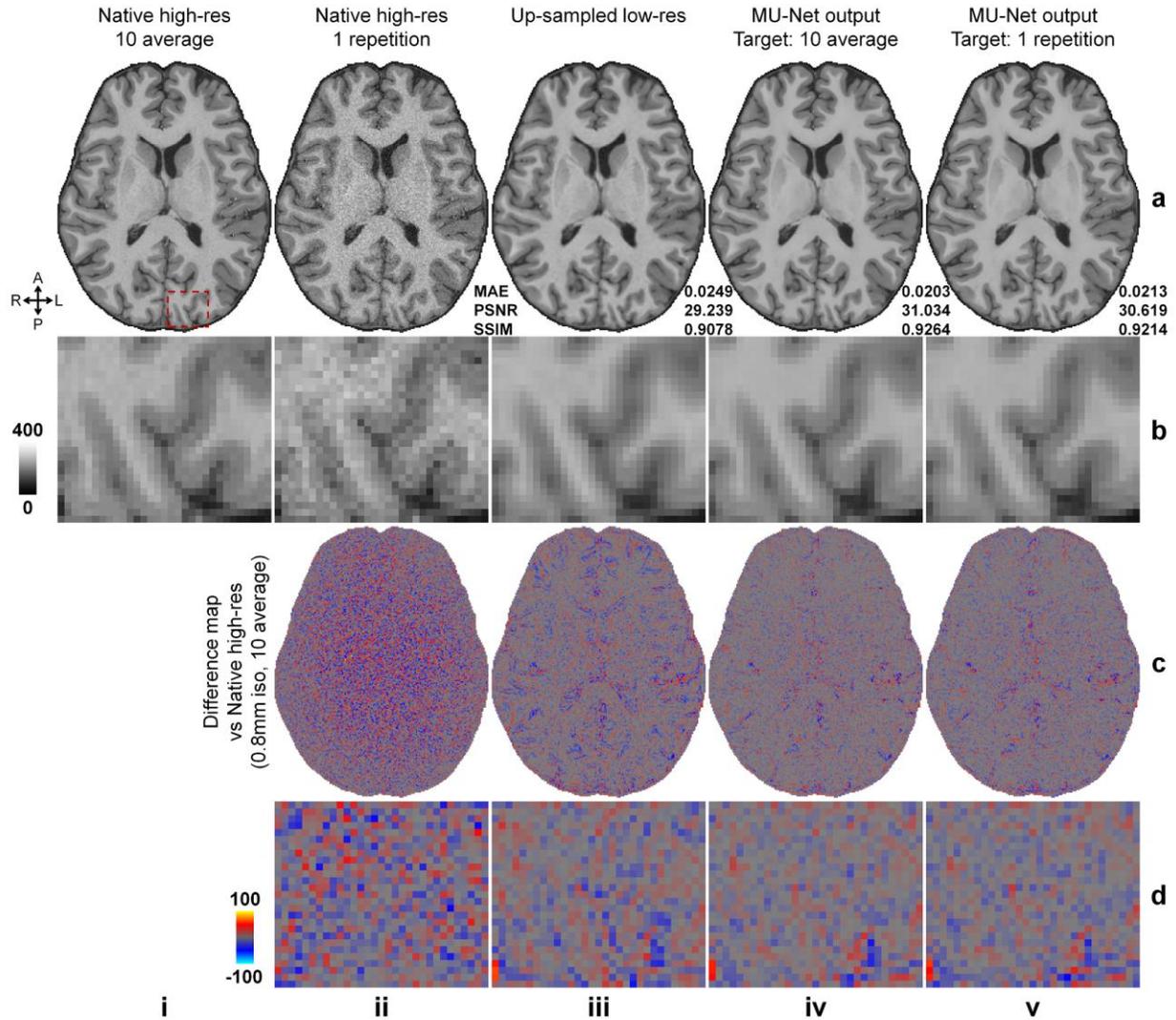

**Figure 4. Empirical image results.** Exemplary axial image slices from 10-repetition averaged high-resolution (a, i), single-repetition high-resolution (a, ii), up-sampled (a, iii), and super-resolved images from NNs trained with 10-repetition averaged and single-repetition images from another 4 subjects are shown with enlarged views (b). The difference maps (c, d) depict the similarity compared to the 10-repetition averaged high-resolution image. Mean absolute error (MAE), peak SNR (PSNR), and structural similarity index (SSIM) are listed to quantify the similarity (a).

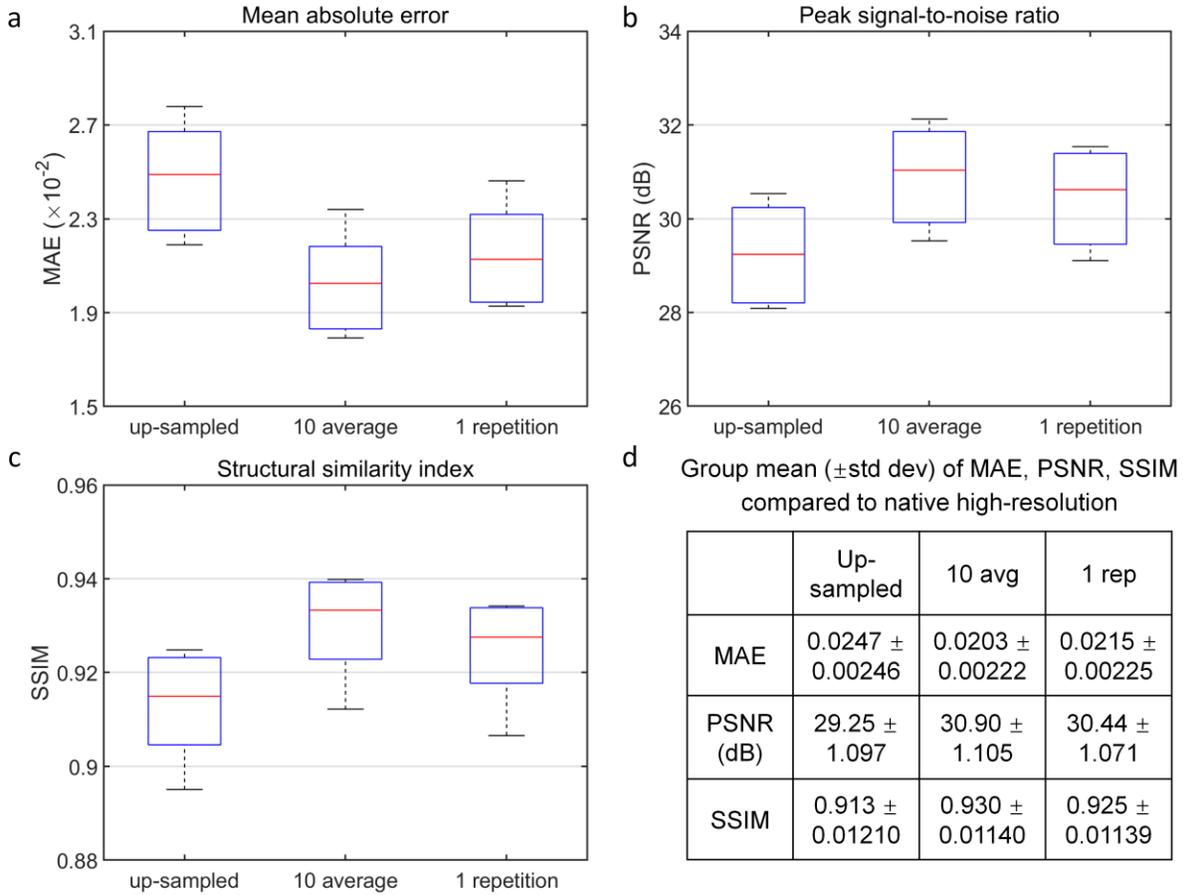

**Figure 5. Empirical image similarity metrics.** Group mean (± standard deviation) of the whole-brain averaged mean absolute error (MAE) (a), peak SNR (PSNR) (b), and structural similarity index (SSIM) (c) of the up-sampled low-resolution images and super-resolution results of neural networks trained with 10-repetition averaged and single-repetition high-resolution images across 5 evaluation subjects. Detailed values are shown in table (d).